\documentclass[aps,prb,twocolumn,amsmath,amssymb]{revtex4-2}
\usepackage{graphicx}
\usepackage{dcolumn}
\usepackage{bm}
\usepackage{amsmath}
\usepackage{color}

\begin{document}

\def\gsim{\mathop {\vtop {\ialign {##\crcr 
$\hfil \displaystyle {>}\hfil $\crcr \noalign {\kern1pt \nointerlineskip } 
$\,\sim$ \crcr \noalign {\kern1pt}}}}\limits}
\def\lsim{\mathop {\vtop {\ialign {##\crcr 
$\hfil \displaystyle {<}\hfil $\crcr \noalign {\kern1pt \nointerlineskip } 
$\,\,\sim$ \crcr \noalign {\kern1pt}}}}\limits}


\title{Crystalline Electronic Field and Magnetic Anisotropy in Dy-based Icosahedral Quasicrystal and Approximant}


\author{Shinji Watanabe$^{*}$ and Tatsuya Iwasaki$^{2}$}
\affiliation{Department of Basic Sciences, Kyushu Institute of Technology, Kitakyushu, Fukuoka 804-8550, Japan \\
$^2$Department of Applied Chemistry, Kyushu Institute of Technology, Kitakyushu, Fukuoka 804-8550, Japan}


\date{\today}

\begin{abstract}
The lack of the theory of the crystalline electric field (CEF) in rare-earth based quasicrystal (QC) and approximant crystal (AC) has prevented us from understanding the electronic states. Recent success of the formulation of the CEF theory on the basis of the point charge model has made it possible to analyze the CEF microscopically. Here, by applying this formulation to the QC Au-SM-Dy (SM=Si, Ge, Al, and Ga) and AC, we theoretically analyze the CEF. In the Dy$^{3+}$ ion with $4f^9$ configuration, the CEF Hamiltonian is diagonalized by the basis set for the total angular momentum $J=15/2$. The ratio of the valences of the screened ligand ions $\alpha=Z_{\rm SM}/Z_{\rm Au}$ plays an important role in characterizing the CEF ground state. For $0\le\alpha<0.30$, the magnetic easy axis for the CEF ground state is shown to be perpendicular to the mirror plane. On the other hand, for $\alpha>0.30$, the magnetic easy axis is shown to be lying in the mirror plane and as $\alpha$ increases, the easy axis rotates to the clockwise direction in the mirror plane at the Dy site and tends to approach the pseudo 5 fold axis. Possible relevance of these results to experiments is discussed.
\end{abstract}


\maketitle


\section{Introduction}
 
The quasicrystal (QC) discovered by D. Shechtman in 1984 has no 
translation 
periodicity 
 and thus may exhibit an orientational 
 rotational symmetry forbidden in periodic crystal~\cite{Shechtman}. 
Clarification of the electric states and the physical properties has been interesting and challenging problem as the frontier of the condensed matter physics because the Bloch theorem based on the periodicity can no longer be applied. 
Recent discoveries of the quantum critical phenomena in the QC Au$_{51}$Al$_{34}$Yb$_{15}$~\cite{Deguchi,Watanuki} and superconductivity Al$_{14.9}$Mg$_{44.1}$Zn$_{41.0}$~\cite{Kamiya} have attracted great interest~\cite{SIDI2022}. 

Since the discovery of the QC, 
it has been a long-standing issue whether the magnetic long-range order is realized in the QC. 
Experimental efforts to find it have been devoted extensively~\cite{Goldman2013}. 
There exists periodic crystal with the local atomic configuration common to that in the QC, which is referred to the approximant crystal (AC). 
In the rare-earth based 1/1 AC Cd$_6$R (R=Pr, Nd, Sm, Gd, Tb, Dy, Ho, Eu, Tm)\cite{Tamura2010,Mori,Tamura2012} and 1/1 AC Au-SM-R (SM=Si, Al, Ge, Sn; R=Gd, Tb, Dy, Ho)\cite{Hiroto2013,Hiroto2014,Das,Sato2019,Hiroto}, magnetic long-range orders have been observed~\cite{Suzuki2021}. 
Recently, ferromagnetic (FM) long-range order has been discovered in the QC Au$_{65}$Ga$_{20}$R$_{15}$ (R=Gd, Tb)~\cite{Tamura}. 

\begin{figure*}[tb]
\includegraphics[width=18cm]{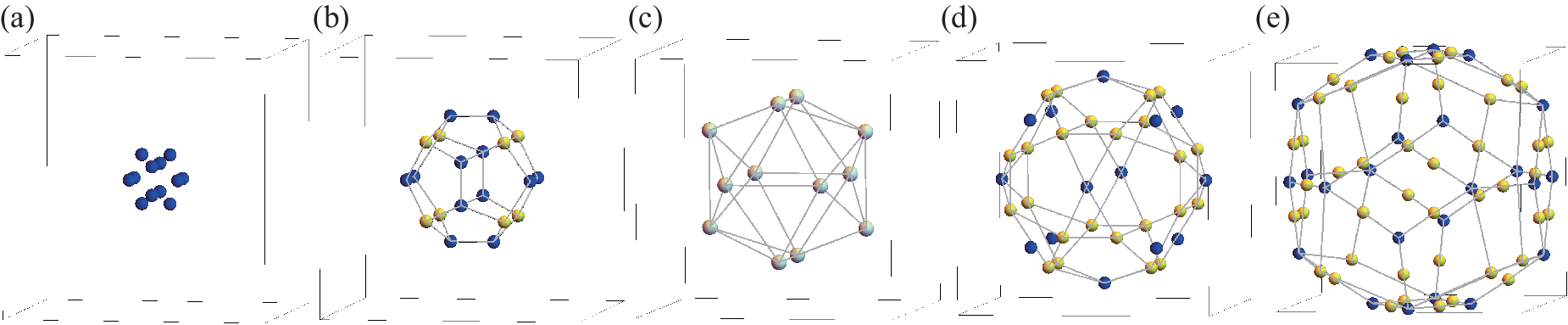}
\caption{(Color online)  
Tsai-type cluster consists of (a) cluster center, (b) dodecahedron, (c) icosahedron (IC), (d) icosidodecahedron, and (e) defect rhombic triacontahedron with Au (yellow), SM (blue), and R (gray). 
The frame is the unit cell of the bcc lattice of the 1/1 AC. 
}
\label{fig:atoms}
\end{figure*}

These intriguing phenomena such as the magnetic long-range orders and the quantum critical phenomena 
take place in the rare-earth based QC and AC Au-SM-R where the 4f electrons at the rare-earth atoms are responsible for the appearance mechanism. 
The QCs and ACs are composed of the Tsai-type cluster which consists of multiple shell structures of atoms. In the 1/1 AC, the Tsai-type cluster is located at the center and corner of the body-centered cubic (bcc) lattice. As a typical example of Au-SM-R, the Tsai-type cluster in the 1/1 AC Au$_{70}$Si$_{17}$Tb$_{13}$~\cite{Hiroto} is shown in Fig.~\ref{fig:atoms}(a)-(e). 
The Tsai-type cluster consists of (a) cluster center, (b) dodecahedron, (c) icosahedron (IC), (d) icosidodecahedron, and (e) defect rhombic triacontahedron. At the 12 vertices of the IC in Fig.~\ref{fig:atoms}(c), the rare earth atom R (gray) is occupied.  
The local configuration of atoms around the rare-earth atom R in the Tsai-type cluster is shown in Fig.~\ref{fig:local_Dy}(a). 

To understand the 4f electronic state in the rare-earth compounds, it has been well known that the crystalline electric field (CEF) is important in periodic crystals~\cite{Stevens,Hutchings}. 
It is also expected to be true in the QC.  
Thus far, experimental studies on the CEF were reported in Tm- and Tb-based 1/1 ACs. 
By the specific-heat and magnetization measurements in Zn$_{85.5}$Sc$_{11}$Tm$_{3.5}$, the CEF energy levels and the eigenstate were discussed~\cite{Jazbec}. 
By the inelastic neutron measurements in Cd$_6$Tb~\cite{Das} and Au$_{70}$Si$_{17}$Tb$_{13}$~\cite{Hiroto}, the CEF energy levels and the eigenstates were analyzed.  In ref.~\cite{Hiroto}, the neutron data were analyzed by the point charge model in detail taking into account the effect of the Au/Si mixed sites. 

Theoretically, the lack of the theory of the CEF prevented us from understanding the 4f-electron state in the rare-earth based QC and AC. 
Recently, the theory of the CEF in the rare-earth based QC and AC has been formulated by the operator equivalence method on the basis of the point charge model~\cite{WK2021}.  
Then, the CEF in the QC Au$_{51}$Al$_{34}$Yb$_{15}$ and 1/1 AC has been analyzed theoretically taking into account the effect of the Au/Al mixed site~\cite{WK2021}. 

Recent theoretical analysis of the CEF in the QC Au-SM-Tb and AC has revealed that the CEF ground state shows the unique magnetic anisotropy~\cite{WSR2021,WPNAS2021}. 
The magnetic anisotropy plays a crucial role in realizing unique magnetic structure on the IC such as the hedgehog, the whirling-moment state, and the ferrimagnetic state~\cite{WSR2021,WPNAS2021}.  
The theoretical calculation based on the magnetic model taking into account the magnetic anisotropy has been performed in the 1/1 AC and the Cd$_{5.7}$Yb-type QC.  
Then, it has succeeded in explaining the magnetic structures of 
not only the measured FM order of the ferrimagnetic state in the 1/1 AC Au$_{70}$Si$_{17}$Tb$_{13}$ and AFM order of the whirling-antiwhirling states in the 1/1 AC Au$_{72}$Al$_{14}$Tb$_{14}$, but also the FM order of ferrimagnetic state in the QC, which offers a candidate for the measured FM order in the  QC Au$_{65}$Ga$_{20}$Tb$_{15}$~\cite{WPNAS2021}. 
 
Very recently, the FM long-range order has been observed in the QC Au-Ga-Dy~\cite{Tamura_pc}. 
Hence, it is important to clarify the CEF in the QC Au-Ga-Dy. 
In this study, by applying the formulation developed in ref.~\cite{WK2021} to  the QC Au-SM-Dy, 
we analyze the CEF at the Dy site in the Tsai-type cluster on the basis of the point charge model. 
We reveal the magnetic easy axis in the CEF ground state, which is indispensable for clarifying the magnetic structure of the magnetically ordered phase observed in the Dy-based QC as well as AC.

\section{Analysis of CEF in QC Au-SM-Dy and AC}

\subsection{Lattice structure}

\begin{figure}[tb]
\includegraphics[width=7cm]{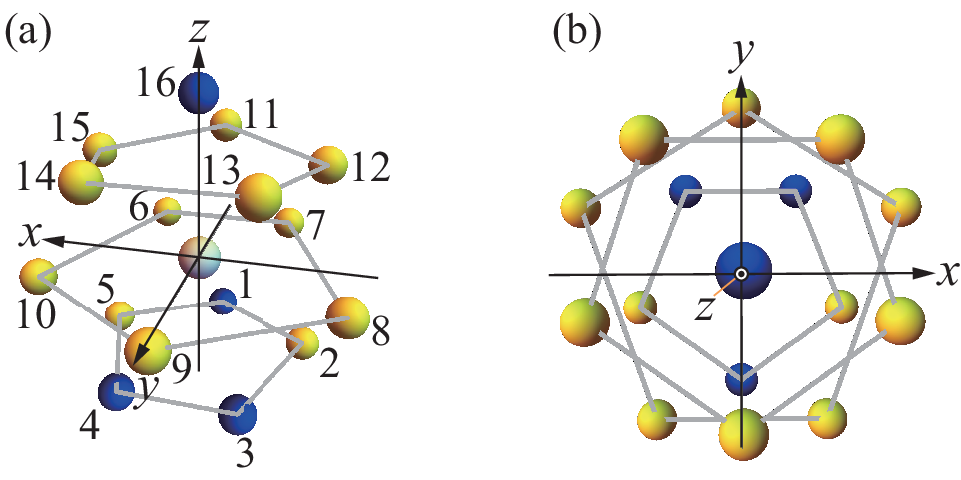}
\caption{(Color online)  
(a) Local configuration of atoms around the rare-earth R atom (gray). 
The number labels surrounding Au (yellow) and SM (blue) (see text). 
(b) Top view of (a). 
}
\label{fig:local_Dy}
\end{figure}

Let us start with the lattice structure. 
Since the atomic coordinate of the QC Au-Ga-Dy and the AC has not been solved experimentally, we employ that in the 1/1 AC Au$_{70}$Si$_{17}$Tb$_{13}$~\cite{Hiroto} shown in Fig.~\ref{fig:atoms} with Tb being replaced with Dy. 
In Fig.~\ref{fig:local_Dy}(a), the vector from the center of the IC passing through the Dy site is set to be the $z$ axis. 
Namely, the $z$ axis is the pseudo 5 fold axis, as clearly seen in the top view shown in Fig.~\ref{fig:local_Dy}(b).  
We set the Dy site as the origin. 
Then, the $y$ axis is set so as the mirror plane to be the $yz$ plane. Finally, the $x$ axis is set so as perpendicular to both the $z$ and $y$ axes. 

In the ternary compounds of the QC Au-SM-Dy and AC, there exists the Au/SM mixed site. Namely, non-rare earth atom Au or SM atom exists at the specific site with some existence probability. For example, in the 1/1 AC Au$_{70}$Si$_{17}$Tb$_{13}$, the $i=1$, 3, 4, and 16th sites are the Au/Si mixed sites with Au/Si: $63\%/37\%$ for the $i=1$, 3, and 4th sites and Au/Si: $19\%/81\%$ for the $i=16$th site in Fig.~\ref{fig:local_Dy}(a)~\cite{Hiroto}.  
In the QC Au-Al-Yb and 1/1 AC, the analysis of the CEF was performed by assuming that the Au/Al mixed site is $100\%$ occupied by Al and the result was compared by the calculation taking into account the effect of the Au/Al mixed site~\cite{WK2021}. The result shows that the former approach describes at least qualitative feature of the CEF. 
The same conclusion was also reported in the QC Au-SM-Tb and AC~\cite{WSR2021,WPNAS2021}. Hence, here we assume that the Au/SM mixed sites are $100\%$ occupied by SM as a first step of analysis.

\subsection{Point charge model for Dy$^{3+}$ ion}

The Dy$^{3+}$ ion has $4f^{9}$ electron configuration. Since the 4f shell becomes closed shell with $4f^{14}$ electron configuration, the Dy$^{3+}$ ion is regarded to have $4f^{5}$ hole configuration. 
Then, the electrostatic interaction between the $4f^{5}$ hole and surrounding ligand ions are expressed as the CEF Hamiltonian 
\begin{eqnarray}
H_{\rm CEF}=5|e|V_{\rm cry}({\bm r}),
\label{eq:H_CEF}
\end{eqnarray}
where the Coulomb potential is given by
\begin{eqnarray}
V_{\rm cry}({\bm r})&=&\sum_{i=1}^{16}\frac{q_i}{\left|{\bm R}_i-{\bm r}\right|}. 
\label{eq:Vcry1}
\end{eqnarray}
Here, ${\bm R}_i$ is the position vector of the Au or SM ions labeled by $i=1\sim 16$ surrounding the Dy$^{3+}$ ion in Fig.~\ref{fig:local_Dy}(a). 
The charge of the ligand ion $q_i$ is given by $q_i=Z_{\rm SM}|e|$ for $i=1,3, 4$ and 16 and $q_i=Z_{\rm Au}|e|$ otherwise, where $Z_{\rm SM}$ and $Z_{\rm Au}$ are the valences of the SM ion and the Au ion, respectively. 
Since the existence probability of the cluster center shown in Fig.~\ref{fig:atoms}(a) is very small~\cite{Hiroto}, here we neglect it for simplicity.

By applying the  Wigner-Eckart theorem~\cite{Wigner} to $H_{\rm CEF}$, we obtain~\cite{WK2021}
\begin{eqnarray}
H_{\rm CEF}= 
\sum_{\ell=2,4,6}\left[B_{\ell}^{0}(c)O_{\ell}^{0}(c)+\sum_{\eta=c,s}\sum_{m=1}^{\ell}B_{\ell}^{m}(\eta)O_{\ell}^{m}(\eta)\right], 
\label{eq:HCEF2}
\end{eqnarray}
where $O_{\ell}^{m}(c)$ and $O_{\ell}^{m}(s)$ are the Stevens operators which consist of the total angular momentum operators 
$J_z$, $J_{+}\equiv(J_x+J_y)/2$, and $J_{-}\equiv(J_x-J_y)/(2i)$~\cite{Stevens,Hutchings}.  
The explicit forms of $O_{\ell}^{m}(c)$ and $O_{\ell}^{m}(s)$ can be found in supplementary material of ref.~\cite{WK2021}.  
The coefficients $B_{\ell}^{m}(\eta)$ are defined by 
\begin{eqnarray}
B_{2}^{m}&=&5|e|C_{2}^{m}\langle r^{2}\rangle\alpha_Jh_2^{m}, 
\label{eq:B2}
\\
B_{4}^{m}&=&5|e|C_{4}^{m}\langle r^{4}\rangle\beta_Jh_4^{m},
\label{eq:B4}
\\
B_{6}^{m}&=&5|e|C_{6}^{m}\langle r^{6}\rangle\gamma_Jh_6^{m},
\label{eq:B6}
\end{eqnarray}
where $C_{\ell}^{m}$ is given by $C_2^0=\sqrt{\pi/5}$, $C_2^1=2\sqrt{6\pi/5}$, $C_2^2=\sqrt{6\pi/5}$, 
$C_4^0=\sqrt{\pi}/12$, $C_4^1=\sqrt{5\pi}/3$, $C_4^2=\sqrt{10\pi}/6$, $C_4^3=\sqrt{140\pi}/6$, $C_4^4=\sqrt{70\pi}/12$,  
$C_6^0=\sqrt{13\pi}/104$, $C_6^1=\sqrt{546\pi}/52$, $C_6^2=4\sqrt{5460\pi}/832$, $C_6^3=\sqrt{5460\pi}/104$, $C_6^4=3\sqrt{182\pi}/104$, $C_6^5=\sqrt{9009\pi}/52$, and $C_6^6=\sqrt{12012\pi}/208$. 
For the expectation values of $r^{2n}$ by the radial part, 
we employ the results of the Dirac-Fock calculation for the Dy$^{3+}$ ion
$\langle r^2\rangle=0.2188~{\rm \AA}^2$, $\langle r^4\rangle=0.1180~{\rm \AA}^4$, and $\langle r^6\rangle=0.1328~{\rm \AA}^6$~\cite{Freeman} . 
The Stevens factors for the Dy$^{3+}$ ion are given by 
$\alpha_J=-2/(3^2\cdot 5\cdot 7) $, $\beta_J=-2^3/(3^3\cdot 5\cdot 7\cdot 11\cdot 13)$, and $\gamma_J=2^2/(3^3\cdot 7\cdot 11^2\cdot 13^2)$~\cite{Stevens,Hutchings}. 

In Eqs~(\ref{eq:B2}), (\ref{eq:B4}), and (\ref{eq:B6}), 
$h_{\ell}^{m}$ is defined by 
\begin{eqnarray}
h_{\ell}^{m}&=&{\rm Re}\left[\sum_{i=1}^{16}\frac{q_i}{R_i^{\ell+1}}(-1)^{m}Y_{\ell}^{m}(\theta_i,\varphi_i)\right].  
\label{eq:hlm}
\end{eqnarray}
In Eq.~(\ref{eq:HCEF2}), $B_{\ell}^{m}(s)$ is obtained by inputting $h_{\ell}^{m}(s)$ into $B_{\ell}^{m}$ instead of 
$h_{\ell}^{m}$, where 
$h_{\ell}^{m}(s)$ is defined by 
\begin{eqnarray}
h_{\ell}^{m}(s)&=&{\rm Im}\left[\sum_{i=1}^{16}\frac{q_i}{R_i^{\ell+1}}(-1)^{m}Y_{\ell}^{m}(\theta_i,\varphi_i)\right].
\label{eq:hlms}
\end{eqnarray}

For the Dy$^{3+}$ ion, the total angular momentum is $L=5$ and the total spin is $S=5/2$. 
Since the $4f^9$ configuration has more than half number of electrons in the closed shell, 
the Hund's rule tells us that the total angular momentum $J=L+S=15/2$ gives the lowest energy state.
Then we calculate $H_{\rm CEF}$ in Eq.~(\ref{eq:HCEF2}) in terms of the basis $|J=15/2, J_z\rangle$ for $J_z=15/2, 13/2, \cdots, -15/2$. 
By diagonalizing the $16\times 16$ matrix of $H_{\rm CEF}$, we obtain the CEF energy $E_n$ $(n=0, 1, 2, \cdots, 7)$ and the eigenstate 
\begin{eqnarray}
|\psi_{n+}\rangle&=&\sum_{J_z=-15/2}^{15/2}a_{n,J_z}\left.\left|J=\frac{15}{2}, J_z\right.\right\rangle, 
\label{eq:ESp}
\\
|\psi_{n-}\rangle&=&\sum_{J_z=-15/2}^{15/2}(-1)^{\frac{2J_z+1}{2}}a_{n,-J_z}^*\left.\left|J=\frac{15}{2}, J_z\right.\right\rangle,
\label{eq:ESm}
\end{eqnarray}
where $a_{n,J_z}$ is the coefficient satisfying the normalization condition $\sum_{J_z=-15/2}^{15/2}|a_{n,J_z}|^2=1$. 
Here, $\pm$ denotes the Kramers states.  
Since the Dy$^{3+}$ ion has the odd-number of electrons (holes), the Kramers degeneracy exists, where the Kramers pair satisfies the orthogonality condition 
$\langle\psi_{n+}|\psi_{n-}\rangle=0$. 

The QC Au-SM-Dy and the AC are metallic crystals. 
The nominal valences of the Au ion and for example Ga ion are known to be +1 and  +3 respectively~\cite{Suzuki2021,Pearson}. However, in reality, these values are considered to be reduced by the effect of the screening of electrons. Hence, we consider the valences of the screened ligand ions $Z_{\rm Au}$ and $Z_{\rm SM}$. In the present study, as a typical value, we set $Z_{\rm Au}=0.223$ which was identified by the inelastic neutron measurement in the 1/1 AC Au$_{70}$Si$_{17}$Tb$_{13}$~\cite{Hiroto} and analyze the CEF by changing the ratio $\alpha=Z_{\rm SM}/Z_{\rm Au}$ as a parameter. 
It is noted here that if the value of $Z_{\rm Au}$ is changed, the CEF energy $E_n$ is changed with $E_n$ being multiplied by a constant factor but the eigenstate does not change as far as $\alpha$ is the same. This point will be discussed again in \S\ref{sec:CE}.

\section{Results} 

\subsection{CEF energy levels}

\begin{figure}[tb]
\includegraphics[width=7cm]{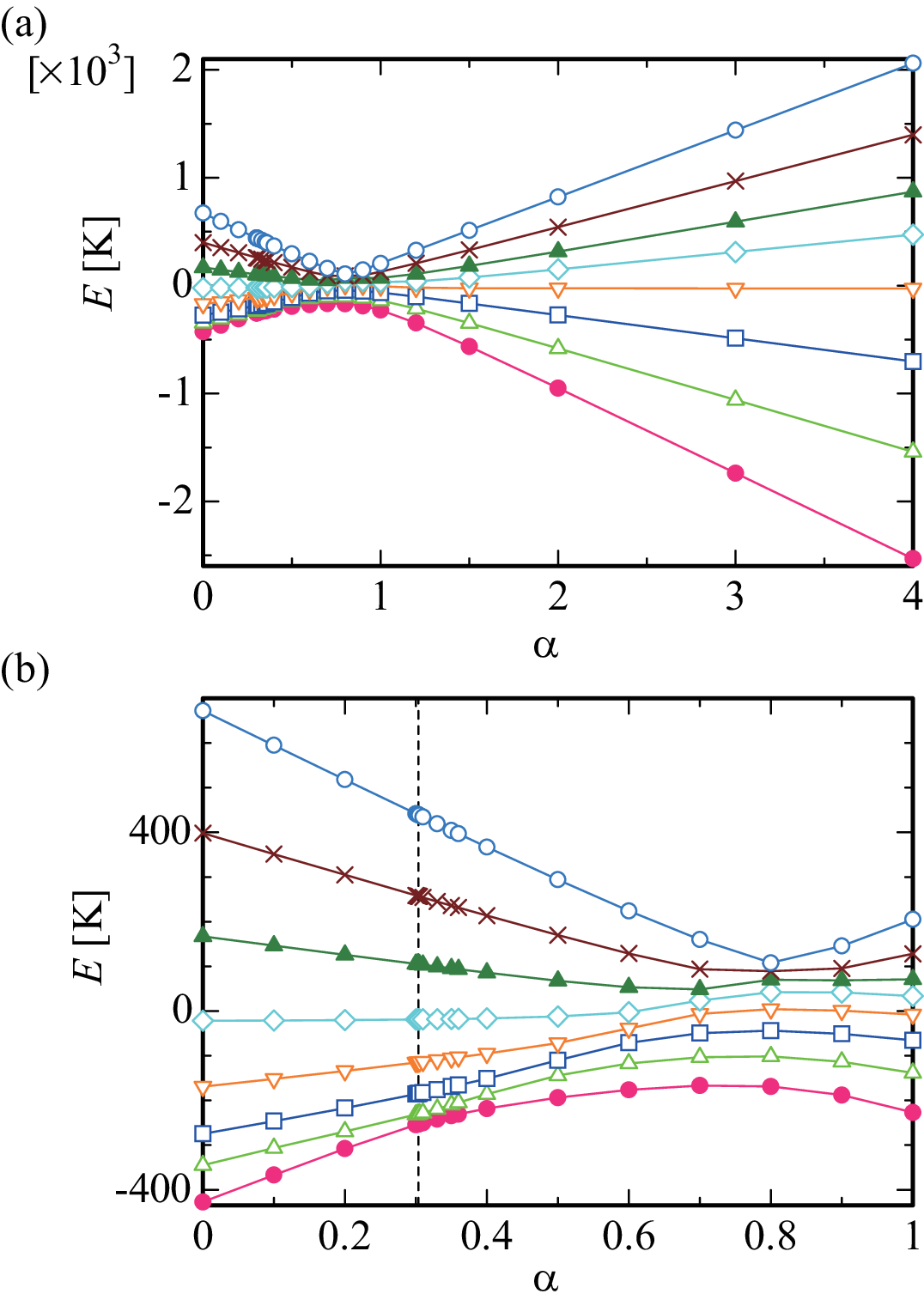}
\caption{(Color online)  
The $\alpha$ dependence of the CEF energy level $E_n$ for $n=0, 1, 2, \cdots, 7$ at $Z_{\rm Au}=0.223$. 
(a) $0\le\alpha\le 4$, (b) $0\le\alpha\le 1$. In (b), the vertical dashed line is plotted at $\alpha=0.3035$. 
}
\label{fig:E_a}
\end{figure}

The $\alpha$ dependence of the CEF energy $E_n$ is shown in Fig.~\ref{fig:E_a}(a). 
Energy levels are split into 8 levels where each level has the Kramers degeneracy. Around $\alpha=1$, the split width $E_7-E_0$ is small,  where the charge distribution around the Dy site is regarded as uniform one. 
In Fig.~\ref{fig:E_a}(b), the enlargement for $0\le\alpha\le 1$ is shown, where the split width of the energy levels is smallest at $\alpha\approx 0.8$. 
At $\alpha=0.3035\equiv\alpha_{\rm c}$ indicated by the vertical dashed line, the CEF ground energy $E_0$ and the first excited energy $E_1$ get close.
Namely, the gap $\Delta=E_{1\pm}-E_{0\pm}$ becomes minimum $\Delta=22.6$~K for $\alpha=0.304$.
Figure~\ref{fig:E_a}(b) shows that $\Delta$ is typically larger than $10\sim 100$~K. Hence, the CEF ground state is expected to dominate over the lower temperature region.

\subsection{Magnetic easy axis in the CEF ground state}

\begin{figure}[tb]
\includegraphics[width=7cm]{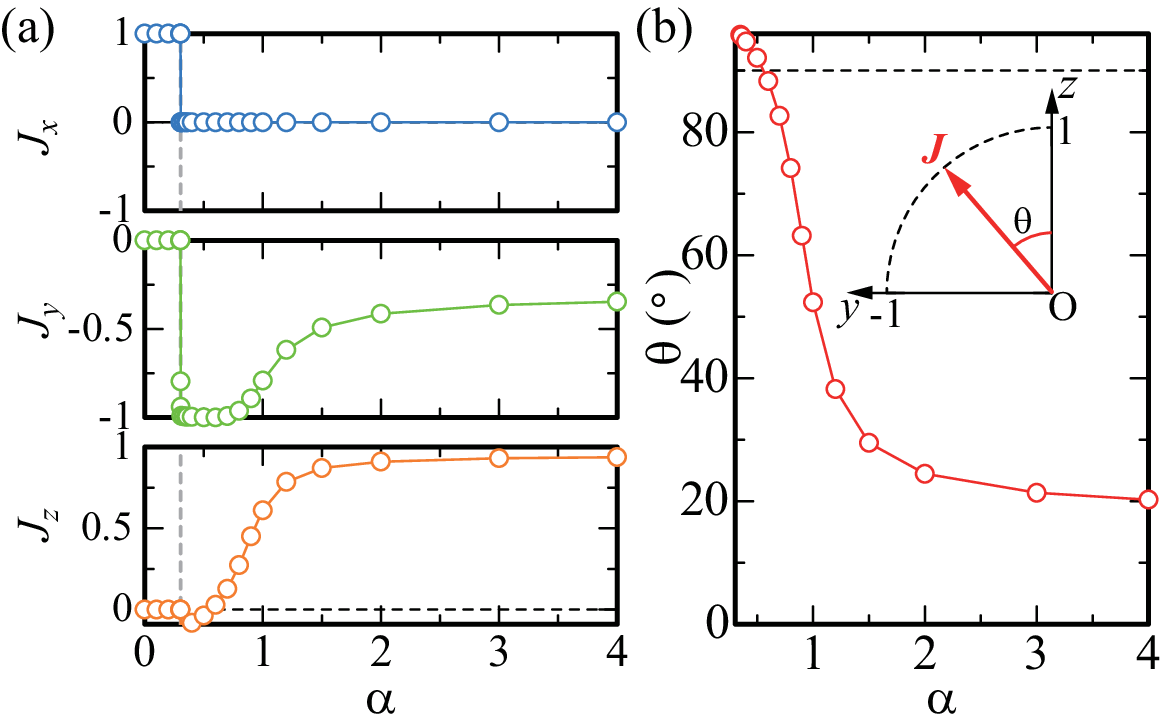}
\caption{(Color online)  
(a) The $\alpha$ dependence of ${\bm J}$ for $Z_{\rm Au}=0.223$.
Top panel: $J_x$ vs. $\alpha$. 
Middle panel: $J_y$ vs. $\alpha$. 
Bottom panel: $J_z$ vs. $\alpha$. 
The vertical dashed line is plotted at $\alpha=0.3035$. 
(b) The $\alpha$ dependence of $\theta$ defined by the angle between ${\bm J}$ and the $z$ axis for $\alpha>0.3035$. 
The horizontal dashed line is plotted at $\theta=90^{\circ}$. 
}
\label{fig:J_a}
\end{figure}

\begin{figure}[tb]
\includegraphics[width=3cm]{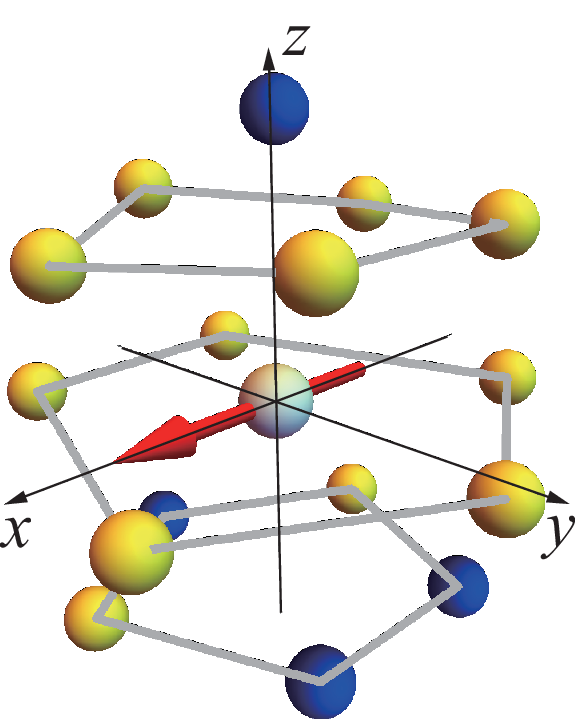}
\caption{(Color online) 
The magnetic easy axis in the CEF ground state for $0\le\alpha<0.3035$.  
The red arrow indicates the ${\bm J}$ vector (see text).  
The local coordinate with Au (yellow) and SM (blue) surrounding Dy (gray) is set in the same way as Fig.~\ref{fig:local_Dy}(a).  
}
\label{fig:J_100}
\end{figure}

To clarify the magnetic anisotropy arising from the CEF, we calculate $M_{\xi\nu}\equiv\langle\psi_{0\eta}|\hat{J}_{\xi}\hat{J}_{\nu}|\psi_{0\eta}\rangle$ $(\xi, \nu=x, y,$ and $z)$. By diagonalizing  $3\times 3$ matrix $M$, we obtain the normalized eigenvector ${\bm J}$ of the largest eigenvalue, which gives the largest magnetic moment direction in the CEF ground state. 
The result is shown in Fig.~\ref{fig:J_a}(a).  
For $\alpha<\alpha_{\rm c}$, the result shows ${\bm J}=(1,0,0)$, which is perpendicular to the mirror $(yz)$ plane.   
Figure~\ref{fig:J_100} illustrates the magnetic easy axis perpendicular to the mirror $(yz)$ plane in the local configuration of atoms surrounding the Dy site where the local coordinate is set in the same way as Fig.~\ref{fig:local_Dy}(a).

On the other hand, 
for $\alpha>\alpha_{\rm c}$, the results in Fig.~\ref{fig:J_a}(a) show ${\bm J}=(0, J_y, J_z)$, which indicates that the magnetic easy axis is lying in the mirror $(yz)$ plane. 
In Fig.~\ref{fig:J_a}(b), we plot the $\alpha$ dependence of the angle $\theta$ defined as the angle between ${\bm J}$ and the pseudo 5 fold axis, i.e., $z$ axis [see inset in Fig.~\ref{fig:J_a}(b)]. 
In the bottom panel of Fig.~\ref{fig:J_a}(a), for $\alpha\gsim 0.30$, $J_z$ is slightly negative and becomes positive as $\alpha$ increases. The negative $J_z$ is not an artifact of the calculation but reflects the fact that the magnetic easy axis ${\bm J}$ points to the $\theta>90^{\circ}$ direction from the 5 fold axis. At $\alpha=0.31$, the anisotropy angle is $\theta=96.7^{\circ}$ and at $\alpha=0.56$ the easy axis ${\bm J}$ is perpendicular to the pseudo 5 fold axis, i.e., $\theta=90^{\circ}$. Namely, for $0.30<\alpha<0.56$, $J_z$ in the bottom panel of Fig.~\ref{fig:J_a}(a) is negative.
As $\alpha$ increases, ${\bm J}$ rotates to the clockwise direction in the mirror $(yz)$ plane and tends to approach the pseudo 5 fold axis.

\begin{figure}[tb]
\includegraphics[width=7cm]{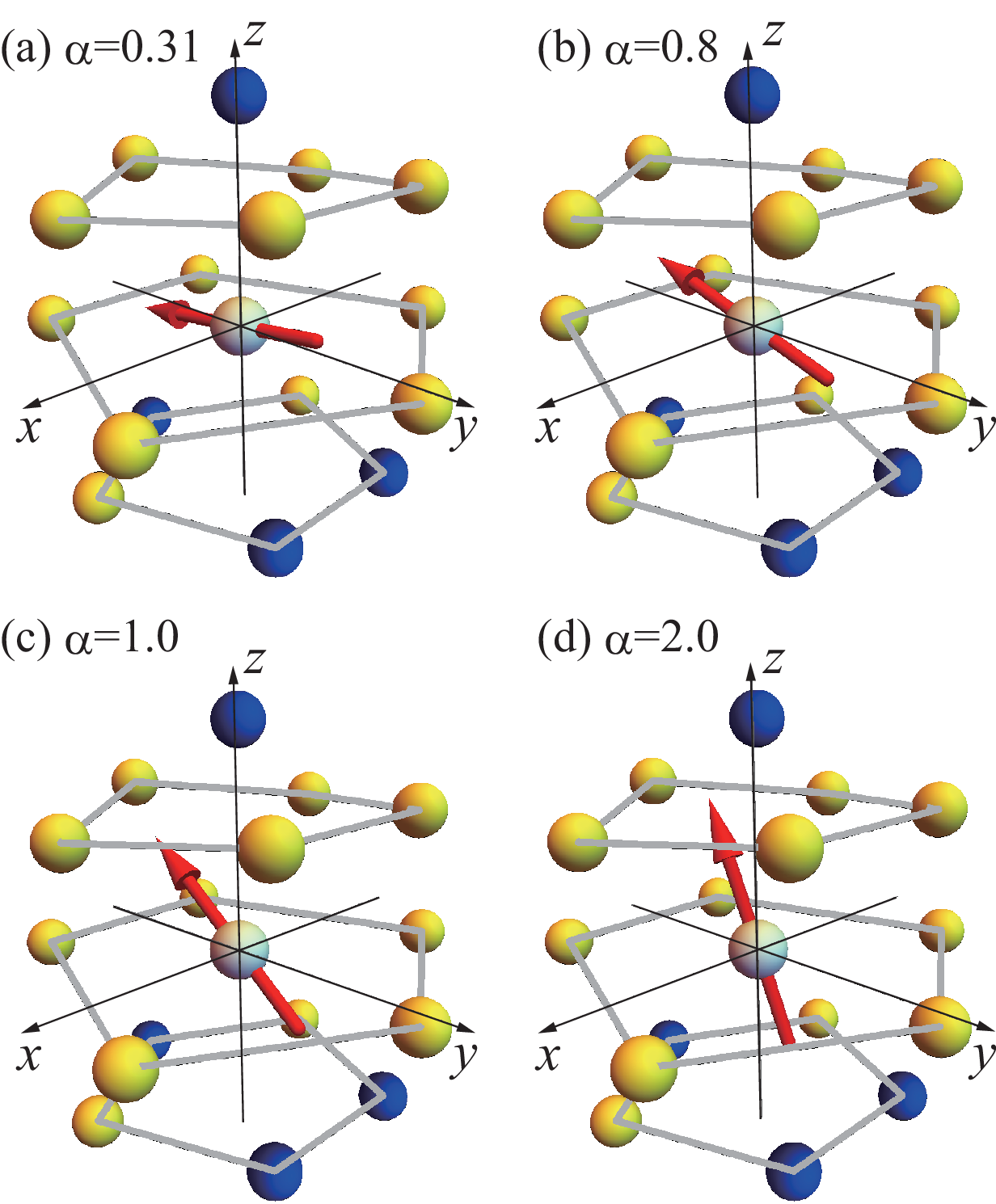}
\caption{(Color online) 
The magnetic easy axis in the CEF ground state for 
(a) $\alpha=0.31$, (b) $\alpha=0.8$, (c) $\alpha=1.0$, and (d) $\alpha=2.0$. 
The red arrow indicates the ${\bm J}$ vector (see text).  
The local coordinate with Au (yellow) and SM (blue) surrounding Dy (gray) is set in the same way as Fig.~\ref{fig:local_Dy}(a).  
}
\label{fig:Dy_mag}
\end{figure}

Figures~\ref{fig:Dy_mag}(a)-(d) illustrate 
the $\alpha$ dependence of the ${\bm J}$ vector, i.e., the magnetic easy axis, in the local configuration of atoms surrounding the Dy site.   
At $\alpha=0.31$, the magnetic easy axis is directed to the nearly perpendicular direction to the pseudo 5 fold axis in the mirror plane, i.e.,  $\theta=96.7^{\circ}$, as shown in Fig.~\ref{fig:Dy_mag}(a).  
At $\alpha=0.8$, $\theta=74.2^{\circ}$ [see Fig.~\ref{fig:Dy_mag}(b)] and hence for $0.31\le\alpha\le0.8$ [see Fig.~\ref{fig:E_a}(b)], the magnetic easy axis remains nearly perpendicular to the mirror plane. 
At $\alpha=1.0$, $\theta=52.3^{\circ}$ [see Fig.~\ref{fig:Dy_mag}(c)]. 
At $\alpha=2.0$, $\theta=24.5^{\circ}$ [see Fig.~\ref{fig:Dy_mag}(d)].

\section{Discussion}

\subsection{Comparison with Au-SM-Tb}

The overall feature of the $\alpha$ dependence of the magnetic easy axis shown in Figs.~\ref{fig:J_a}(a) and \ref{fig:J_a}(b) is similar to that obtained for the CEF analysis of the QC Au-SM-Tb where the Au/SM mixed site was assumed to be $100\%$ occupied by SM~\cite{WPNAS2021}. 
Although Tb$^{3+}$ ion has $4f^{8}$ electron configuration giving non-Kramers system and the ground total angular momentum is $J=6$, 
the CEF ground state was shown to be doubly degenerate. The magnetic easy axis is perpendicular to the mirror plane for small $\alpha$ and is lying in the mirror plane for large $\alpha$ as similar to Fig.~\ref{fig:J_100} and Fig.~\ref{fig:Dy_mag}. 

The magnetic model taking into account the uniaxial anisotropy shown in Fig.~\ref{fig:J_a}(b) 
\begin{eqnarray}
H=-\sum_{\langle i,j\rangle}J_{ij}\hat{\bm J}_i\cdot\hat{\bm J}_j
\label{eq:H_mag}
\end{eqnarray}
was introduced on the IC, 1/1 AC, and the QC~\cite{WSR2021,WPNAS2021}. 
Here, $\hat{\bm J}$ is the unit vector operator whose direction is restricted to either parallel or antiparallel to the moment direction shown in Fig.~\ref{fig:J_a}(b).  
Hence, the phase diagram of the model (\ref{eq:H_mag}) is also applied to the QC Au-SM-Dy and the 1/1 AC. 
For FM interaction $J_{ij}$, the magnetic states on the IC, 1/1 AC, and Cd$_{5.7}$Yb-type QC was shown in ref.~\cite{WPNAS2021}. 
For antiferromagnetic interaction,  $J_{ij}$, the magnetic states on the IC, 1/1 AC, and Cd$_{5.7}$Yb-type QC was shown in ref.~\cite{WSR2021}. 
Present study has revealed that the minimal model for the QC Au-SM-Dy and the AC is the model (\ref{eq:H_mag}), which is also the model for the QC Au-SM-Tb and the AC. 

It is also noted that the magnetic anisotropy shown in Fig.~\ref{fig:J_100} has neither been observed experimentally in Au-SM-Dy nor Au-SM-Tb to date. 
The magnetic structure of the model (\ref{eq:H_mag}) applied to the IC was discussed in supplementary information in ref.~\cite{WPNAS2021}. 
The observation of this state is left for interesting future experiments.

\subsection{Comparison with experiments}
\label{sec:CE}

The results shown in Fig.~\ref{fig:E_a} and Fig.~\ref{fig:J_a} indicate that the ratio of the screened ligand ions $\alpha$ plays an important role in characterizing the CEF ground state. Experimentally, by performing the inelastic neutron measurement, the valences of the screened ligand ions are expected to be determined in the following way.
First, by comparing the ratio of each difference of the energy levels $E_{n+1}-E_n$ for $0\le n\le6$ with Fig.~\ref{fig:E_a}, the parameter $\alpha$ is identified.  Next, to reproduce the absolute value of each energy level $E_n$, a constant factor is multiplied to each $E_n$. Then, the valence of the screened Au ion $Z_{\rm Au}$ is identified by multiplying $Z_{\rm Au}$ by the factor. 
In this way, $Z_{\rm SM}$ as well as $Z_{\rm Au}$ can be determined experimentally.

\section{Summary}

The CEF of the 4f electrons at the Dy site inside the Tsai-type cluster in the QC Au-SM-Dy and AC is analyzed theoretically on the basis of the point charge model. 
The ratio of the valences of the screened ligand ions $\alpha=Z_{\rm SM}/Z_{\rm Au}$ plays an important role in characterizing the CEF ground state. 
For $0\le\alpha<0.30$, the magnetic easy axis for the CEF ground state is perpendicular to the mirror plane. On the other hand, for $\alpha>0.30$, the magnetic easy axis is lying in the mirror plane and as $\alpha$ increases, the easy axis rotates to the clockwise direction in the mirror plane  at the Dy site and tends to approach the pseudo 5 fold axis. 
These overall features are similar to those obtained in the CEF analysis in the QC Au-SM-Tb and AC in ref.~\cite{WPNAS2021} and hence the magnetic model (\ref{eq:H_mag}) taking into account the uniaxial magnetic anisotropy arising from the CEF introduced in refs.~\cite{WPNAS2021,WSR2021} is considered to be also relevant to the QC Au-SM-Dy and AC.  

Experimental identification of $\alpha$ as well as $Z_{\rm Au}$ is important for clarification of the uniaxial anisotropy of the 4f magnetic moment toward the understanding of the magnetic structure of the FM long-range order in the QC Au-Ga-Dy.  
The experimental search for the magnetic structure with uniaxial anisotropy perpendicular to the mirror plane, which is predicted to exist in the present study, is also interesting future subject.

\begin{acknowledgments}


S.W. thanks R. Tamura, F. Labib, and S. Suzuki for informing him of experimental data with enlightening discussions.  
This work was supported by JSPS KAKENHI Grants Numbers JP19H00648, JP22H0459, JP22H01170, and JP23K17672. 

\end{acknowledgments}

\end{document}